\documentstyle{article}[12pt]
 \def\be{\begin{equation}}
\def\ee{\end{equation}}
\def\bea{\begin{eqnarray}}
\def\eea{\end{eqnarray}}

\def\parsl{\partial \!\!\! /}

\newcommand{\Section}[1]{\section{#1}\setcounter{equation}{0}}

\def\nn{\nonumber}

\def\p{\partial}

\def\l{\label}
\def\laa4{\lambda _{\psi\p`i\psi\psi}}
\def\laap4{\lambda _{\psi\psi\lambda\lambda}}
\def\la3{\lambda _{\psi\phi\lambda}}

\def\){\right)}
\def\({\left(}
\def\D{\delta}

\begin{document}
\begin{titlepage}
\begin{flushright}
\end{flushright}

\pagestyle{plain}
\vskip .05in
\begin{center}
\Large{\bf
Quantizing field theories in noncommutative geometry
and the correspondence between anti de Sitter
space and conformal field theory
}
\small
\vskip .15in

Kamran Kaviani, Amir Masoud Ghezelbash
\footnote{{\sl e-mail:kaviani,amasoud@theory.ipm.ac.ir}
}
\vspace{.5 cm}
\small

{\it Department of Physics, Az-zahra University,
, Tehran 19834, Iran}\\

\vspace{.3 cm}

{\it Institute for Studies in Theoretical Physics and Mathematics (IPM),}\\
{\it P.O.Box 19395-5531, Tehran, Iran}\\

\vspace{.3 cm}


\begin{abstract}
By using the approach of non-commutative geometry, we study spinors and scalars
on the two layers AdS$_{d+1}$ space. We have found that in the boundary of
two layers AdS$_{d+1}$ space, by using the AdS/CFT correspondence, we
have a logarithmic conformal field theory. This observation propose a way
to get the quantum field theory in the context of non-commutative geometry.
\end{abstract}
\end{center}
\end{titlepage}
\newpage
\Section{Introduction}

Various aspects of the correspondence between field theories in
$(d+1)$--dimensional Anti de  Sitter space (AdS) and $d$--dimensional
conformal field theories (CFT's) has been studied in the last few months. An
important example is the conjectured correspondence between the large $N$
limits of certain conformal field theories in $d$--dimensions and
supergravity on the product of a $(d+1)$--dimensional AdS space with a compact
manifold \cite{Ma}. This suggested correspondence was made more precise in
\cite{GKP,W}.

The general correspondence between a theory on an AdS and a conformal theory
on the boundary of AdS is the following. Consider the partition function of
a field theory on AdS, subjected to the constraint
\be\label{1}
\phi_{\vert_{\partial AdS}}=\phi_0,
\ee
that is
$
Z_{\rm AdS}[\phi_0]=\int_{\phi_0}{\rm D}\phi\exp \{iS[\phi ]\},
$
where the functional integration is over configurations satisfying
(\ref{1}). It is well known that the symmetry algebra of a
$(d+1)$--dimensional
AdS is O($d$,2), which is the same as the conformal algebra on a $d$--
dimensional Minkowski space. From this, it is seen that
$Z_{\rm AdS}[\phi_0]$ is invariant under conformal transformations, and this
is the root of the analogy between theories on AdS and conformal theories on
$\partial{\rm AdS}$. In fact, if
$
\phi\to O\phi,
$
is a space--time symmetry of the theory on AdS, it is seen that
$
Z_{\rm AdS}[O^{-1}\phi_0]=\int_{\phi_0}{\rm D}(O\phi )\exp \{iS[O\phi ]\},
=\int_{\phi_0}{\rm D}\phi\exp \{iS[\phi ]\},
$
which means that
$
Z_{\rm AdS}[O^{-1}\phi_0]=Z_{\rm AdS}[\phi_0].
$
So, one can use $Z_{\rm AdS}[\phi_0]$ as the generating function of a
conformally invariant theory on the boundary of AdS, with $\phi_0$ as the
current.

The length element in AdS$_{d+1}$ takes the form
\be\label{10}
{\rm d}s^2={{-({\rm d}x^0)^2+\sum_{i=1}^d({\rm d}x^i)^2}\over{(x^d)^2}}.
\ee
and
almost all of the boundary is now contained in $x^d=0$.

The above mentioned correspondence have been studied for various cases,
e.g. a free massive scalar field and a free U(1) gauge theory \cite{W}, an
interacting massive scalar field theory \cite{MV}, free massive spinor
field theory \cite{HS}, and interacting massive spinor-scalar field theory
\cite{GKPF}. Also, the group-theoretic interprtation of correspondence
has been studied in \cite{DOB}.
Our aim in this article is to shed light, in one hand, to the
correspondence between theories on the non-commutative AdS spaces and
logarithmic conformal field theories (LCFT's),
and on the other hand, to the
construction of quantum field theory in the non-commutative geometry.
In this context, we give a geometrical interpretation of the derived
theories which are considered in \cite{GHEZ}.
In this article, we will build the
toy models for
spinors and scalars in AdS$_{d+1}$ space by using the generalized Dirac operator
and we
will show that on the boundary of the AdS$_{d+1}$, there will be a logarithmic
conformal quantum field theory.
In section 2 and 3, we review in brief, the LCFT's and non-commutative
geometry. In section 4 and 5, we study the spinor and scalar fields
in two layers AdS space respectively.
\Section{A brief review of logarithmic conformal field theories}

It has been shown by Gurarie \cite{Gu}, that conformal field
theories whose correlation functions exhibit logarithmic behaviour,
can be consistently defined. It is shown that if in the OPE of two local
fields, there exist at least two fields with the same conformal dimension,
one may find some special operators, known as logarithmic
operators. As discussed in \cite{Gu}, these operators with the ordinary
operators form the basis of a Jordan cell for the $L_i$'s (the
generators of the conformal algebra).
In some interesting physical theories,
one can naturally find
logarithmic terms in the correlators of theories.
\cite{Sa},
Logarithmic conformal field theories for
$d$--dimensional case ($d>2$) has also been studied \cite{GhK}.
The basic properties of logarithmic operators are that
they form a part of the basis of the Jordan cell
for $L_i$'s, generators of the Virasoro algebra, and in the correlator of
such fields there  is a logarithmic singularity \cite{Gu}.
In \cite{RAK} and \cite{RAKK} assuming conformal invariance  two-- and
three--point functions for the case of one or more logarithmic fields
in a block, and one or more  sets of logarithmic fields
have been explicitly calculated. Regarding logarithmic fields
{\it formally} as derivatives of ordinary fields with respect to their
conformal dimension,  $n$--point functions containing logarithmic fields
have been calculated in terms of those of ordinary fields.
These have been done when conformal weights belong to a discrete set.
In \cite{KAR}, logarithmic conformal field theories with continuous weights
have been considered. It is shown in \cite{RAK} and \cite{RAKK} that
if the set of weights
is discrete, when the Jordan cell for $L_i$ is two dimensional,
there are two fields ${\cal O}$ and ${\cal O}'$, with the following
two-point functions
\be \l{T}
   \langle{\cal O}({\bf x}){\cal O}({\bf y}) \rangle =0,
\ee
\be \l{TT}
   \langle{\cal O}({\bf x}){\cal O}'({\bf y}) \rangle =
   \frac{c}{|{\bf x}-{\bf y}|^{2\Delta}},
\ee
\be \l{TTT}
   \langle{\cal O}'({\bf x}){\cal O}'({\bf y}) \rangle =
   \frac{1}{|{\bf x}-{\bf y}|^{2\Delta}}
   \left( c'-2c\ln |{\bf x}-{\bf y}|
   \right).
\ee

In \cite{KA} another type of derivation is also introduced. The main idea of
this construction is based on the {\it formal} derivation of the entities of
the original system with respect to a parameter, which may or may not
explicitly appear in the original theory. One way of viewing this is through
the concept of contraction: consider two systems with parameters $\lambda$ and
$\lambda +\D$. These two systems are independent to each other. One can write an
action as the difference of the actions of the two system divided by $\D$ to
describe both systems. One can use one of these degrees of freedom and the
difference of them divided by $\D$ a new set of variables. This system,
however, is equivalent to two copies of the original system. But if one lets
$\D$ tend to zero, a well--defined theory of double number of variables is
obtained, which no longer can be decomposed to two independent parts. One
can, however, solve this theory in terms of the solution of the original
theory. This procedure is nothing but a contraction. It has been shown in
\cite{KAA} that any symmetry, and any constant of motion of the original
theory, results in a symmetry and a constant of motion of the derived one
and  any theory derived from an integrable theory is integrable. At last, it
has also been shown that this technique is applicable to classical field
theories as well. This technique is applicable to quantum systems as well.
Here, however, a novel property arises: the derived quantum theory is
{\it almost classical}; that is, in the derived theory there are only
one--loop quantum corrections to the classical action \cite{KAA}. Using this
property, one can calculate all of the Green functions of the derived theory
exactly, even though this may be not the case for the original theory.
\newline
Using the AdS/CFT correspondence, in \cite{GHEZ}, a
correspondence between  field theories in $(d+1)$--dimensional AdS space and
$d$--dimensional logarithmic conformal field theories has been obtained.
It has been showed
that by a suitable choice of action in AdS$_{d+1}$,
one gets a LCFT on the $\partial$AdS. In general, using any field theory on
AdS, which corresponds to a CFT on $\partial$AdS, one can systematically
construct other theories on AdS corresponding to LCFT's on $\partial$AdS.
\newline
\Section{A brief review of non-commutative geometry}

In 1989, Alain Connes and John Lott obtained the lagrangian of the standard
model by an algebraic-geometrical approach which is called non-commutative
geometry or spectral geometry \cite{BISSE}. They considered a two layers
Minkowskian space-time
, which each layer endowed by a suitable bundle for describing the $U(1)
\times SU(2)$ gauge symmetry. What they really has been obtained was a
generalized
Yang-Mills theory in which the Higgs potential appears naturally. This
Higgs field seems to be the component of the generalized gauge field in
the direction of
discretnes of space. But there was some difficulties in their approach such as
their fermions Hilbert space does not match with the phenomenological
standard model lagrangian \cite{BISSH} and another problem was due to the
existance of some relations
between some parameters of the theory such as the mass of the Higgs and
Top quark \cite{BISHAFT}. These relations are not consistent with the flow of
renormalization
of this parameters \cite{BISHASHT}.
The reason for the latter problem is said to be due to taking the incorrect
geometrical space which is a two layers commutative Minkowski space
(or in other words the classical space). By
commutative space, we mean that the algebra which describes the space-time is
$C^\infty (M)$ which is the commutative $C^*$-algebra. Now if one modifies this
algebra to some non-commutative or quantum algebra, he or she can derives
standard model which is in accordance with phenomenological standard model
with the same number of free parameters. \newline
In non-commutative geometry, we express the topological space (which is assumed
to be compact) by a unital $C^*$-algebra ${\cal A}$ which is commutative in the
case of ordinary manifolds. Then for endowing this algebra with a differential
structure, we need the K-cycle $({\cal A},{\cal H},{\cal D})$, where ${\cal H}$
is the
Hilbert space for representating the elements of ${\cal A}$ as the linear
operators and ${\cal D}$ is called the generalized Dirac operator. The role of
K-cycle for non-commutative geometry is similar to the role of differential
structure in ordinary differential geometry. Having a K-cycle, one can develop
a differential algebra $\Omega ({\cal A})$ for the non-commutative space which
is equivalent to differential geometry for the manifolds.

In
correspondence to ordinary lagrangian for spin $\frac{1}{2}$ particles, one
can take the
following statement as the generalized spin $\frac{1}{2}$
action in non-commutative
geometry,
\be \l{seven}
{\cal I}=<\psi\mid{\cal D}\mid\psi>=\int d^{d+1}x\sqrt g\bar \psi {\cal D}\psi.
\ee
where $\psi$ is the spinor field in ${\cal H}$.
In fact, one can take the Dirac operator ${\cal D}$ such that the suitable mass
term emerges naturally in the above statement. In the case of spin $0$
particles in accordance to Klein-Gordon action, one can define the
non-commutative geometric action as follows,
\be \l{eight}
{\cal S}=\frac{1}{4}Tr_{\omega}\{[{\cal D},\phi]\Gamma[{\cal D},\phi]+M^2\phi
\Gamma
\phi\}, \ee
where $\phi$ is an element of the algebra ${\cal A}$, and by $Tr_{\omega}$, we
mean
the Dixmier trace \cite{BISCH}. In the above statement, $\Gamma$
is some gradation
operator.
\Section{Two layers AdS space and spin $\frac{1}{2}$ lagrangian}

At this stage, let us take a two layers space as our topological space.
A suitable algebra which discribes it, is
$ \l{nine}
{\cal A}=C^\infty(AdS)\otimes {\bf M}_2(C),
$
where by $C^\infty(AdS)$, we mean the algebra of complex smooth
functions
defined on AdS, and by ${\bf M}_2(C)$, we mean the algebra of $2\times 2$
complex matrices. Then as a representation for the elements of ${\cal A}$, we
choose the following representation,
$ \l{ten}
\pi (a)=\pmatrix{f_1(x)&0\cr 0&f_2(x)};\quad a\in  {\cal A}.
$
In fact, this representation among the others has a simple interpretation,
simply $f_1(x)$ and $f_2(x)$ can be interpreted as functions on each AdS
spaces. Here we can split our Hilbert space ${\cal H}$ into two Hilbert spaces
${\cal H}_1$ and ${\cal H}_2$, which each of them corresponds to one AdS
layer. So it is logical to define a gradation operator
$\Gamma =\pmatrix{1&0\cr 0&-1}$ for showing the splitting of the Hilbert space
${\cal H}$, and we can represent our spinors as
$\pmatrix{\psi _1 \cr \psi _2}$
. Now, it is the time to
define our Dirac operator to complete the definition of our K-cycle. We take it
as follows,
\be \l{eleven}
{\cal D}=\pmatrix{\parsl +m&M\cr M&-(\parsl +m)}; \quad m,M\in C\otimes 1_{d
\times d}.
\ee
So according to Eq. (\ref{seven}) for spin $\frac{1}{2}$ lagrangian, we have,
\bea \l{twelve}
{\cal L}=\bar \psi {\cal D}\psi&=&(\bar \psi _1 ,\bar \psi _2)
\pmatrix{\parsl -m&M\cr M&-(\parsl +m)}
\pmatrix{\psi _1 \cr \psi _2}
\nonumber\\
&=&\bar \psi _1(\parsl +m)\psi _1+\bar \psi _1M\psi _2-\bar \psi _2(\parsl +m)
\psi _2+\bar \psi _2M\psi _1.
\eea
It seems to be logical to take the physics of both AdS spaces identical,
which means $\psi _1$ and $\psi _2$ to be the same. So at this stage, we
take $\psi _1$ and $\psi _2$ as follows,
$ \l{therteen}
\psi _1=\frac{1}{2}(\psi _++\epsilon \psi _-),\quad
\psi _2=\frac{1}{2}(\psi _+-\epsilon \psi _-),
$
where $\epsilon$ is an infinitesimal number. Now in terms of these new spinors
, the lagrangian can be written as follows,
\be \l{fourteen}
{\cal L}=\bar \psi _+(\parsl +m)\psi _-+\bar \psi _-
(\parsl +m)\psi _++\bar \psi _+M\psi _+,
\ee
which we have redefined the fields $\sqrt {\frac{\epsilon}{2}}\psi _{\pm}$ as
$\psi _{\pm}$. This is exactly the fermionic singleton lagrangian.
However, as observed in \cite{HS}, to obtain the non-vanishing on-shell
spinor action in the usual AdS space, one must  add to the massive free
spinor action a boundary term. For such a boundary term in the
non-commutative geometry, we take the following boundary lagrangian,
\be \l{sup}
{\cal L} _{boundary}=\bar \psi _{boundary} \Gamma
\psi _{boundary},
\ee
where $\psi _{boundary}$ is the projection of $\psi $ on the
boundary of AdS$_{d+1}$.
Hence the total lagrangian of spinors system on the two layers AdS space is,
\bea
{\cal L}_{tot}&=&
\bar \psi _+(\parsl +m)\psi _-+\bar \psi _-
(\parsl +m)\psi _++\bar \psi _+M\psi _+\nn\\ \l{fer}
&+&\{\bar \psi _{+(boundary)}
\psi _{-(boundary)}+\bar \psi _{-(boundary)}\psi _{+(boundary)}\}\delta (x^d).
\eea
By using the same method of Ref. \cite{GHEZ},
one can show that the projection of the fields $\psi _-$
and $\psi _+$ to the boundary are currents for the pseudo-conformal operators
$\lambda _+^\alpha$ and $\lambda _-^\alpha$ respectively. The two point
correlation function
$<\bar \lambda _-^\alpha (\bf{x})\lambda _-^\beta (\bf{y})>$
 vanishes. The two point correlation function
$<\bar \lambda _-^\alpha (\bf{x})\lambda _+^\beta (\bf{y})>$
has a scaling behaviour and
$<\bar \lambda _+^\alpha (\bf{x})\lambda _+^\beta (\bf{y})>$
has a logarithmic behaviour.
\Section{Two layers AdS space and scalar field lagrangian}

In this section, we consider the action (\ref{eight}) which we have introduced
it for the scalar field as the generalization of massive Klein-Gordon
action in the non-commutative geometry.  For the two layers AdS space, we
take the Dirac operator as same as (\ref{eleven}). As it was mentioned in
section
3, the scalar fields are elements of the algebra ${\cal A}$, and again we take
the following representation for the elements of ${\cal A}$ in Hilbert space
${\cal H}$,
$ \l{fifteen}
\pi (\phi)=\pmatrix{\phi _1(x)&0\cr0&\phi _2(x)},
$
where $\phi _1(x)$ and $\phi _2(x)$ are functions in $C^\infty (AdS)$.
By using $\Gamma=\pmatrix{1&0\cr0&-1}$ as the gradation operator, it would
be straightforward to obtain,
\bea
{\cal S}&=&\frac{1}{4}\int d^{d+1}x \sqrt{g}tr\{Tr \big( [{\cal D},\pi (\phi)]
\Gamma
[{\cal D},\pi (\phi)]+m^2 \phi \Gamma \phi \big)\} \nn  \\
\l{sixten}
&=&\frac{1}{4}\int d^{d+1}x \sqrt{g}tr \big( \parsl \phi _1\parsl\phi _1+2M^2
(\phi _2-\phi
_1)^2-\parsl \phi _2\parsl \phi _2+m^2(\phi _1^2-\phi _2^2)\big),
\eea
where $tr$ and $Tr$ are traces on the Clifford and matrix structure
respectively.
We notice that in the first line of eq. (\ref{sixten}), the meaning of
$Tr$ is integrating
over the discrete dimension of space. So, the resultant relation
(\ref{sixten}) could be interpreted as the effective action in the ordinary
AdS$_{d+1}$ space.
By introducing the new fields $\phi _+$ and $\phi _-$ as
follows,
$ \l{seventeen}
\phi _1=\epsilon \phi _++\phi _-,\quad
\phi _2=\epsilon \phi _+-\phi _-,
$
and substituting them in (\ref{sixten}), and preserving the terms up to the
first order in $\epsilon$, we will obtain,
\be
\l{eighteen}
{\cal S}=\frac{1}{4}\int d^{d+1}x \sqrt{g}tr\{4\epsilon \parsl \phi _+\parsl
\phi _-+
2M^2\phi _-^2+4m\epsilon\phi _+\phi _-\}.
\ee
Now, if one redefines $\sqrt{d+1}\epsilon \phi _+$ as $\phi _+$ and
$\sqrt{d+1}\phi _-$ as $\phi _-$, then the last relation becomes,
\be \l{ninteen}
{\cal S}=\int d^{d+1}x \sqrt{g}\{\parsl \phi _+\parsl \phi _-+
\frac{M^2\phi _-^2}{2}+m\phi _+\phi _-\}.
\ee
This action in the literature is known as bosonic singleton action.
In the
Ref. \cite{GHEZ}, by an explicit calculation, it has been showed that the
projection fields $\phi _{-(0)}$ and $\phi _{+(0)}$ obtained through
the fields $\phi _-$ and $\phi _+$ by restricting them to the boundary
of the AdS$_{d+1}$, are the currents for the conformal operators,
${\cal O}_+$ and ${\cal O}_-$ respectively. In fact, the two--point
correlation functions of the operators ${\cal O}_+$ and ${\cal O}_-$ which
live in the boundary of AdS$_{d+1}$ are in the following form;
The correlation function of ${\cal O}_-({\bf x})$ and
${\cal O}_-({\bf y})$ vanishes as the relation (\ref{T}). The
correlation function of
${\cal O}_-({\bf x})$ and ${\cal O}_+({\bf y})$ is in the scaling form
(\ref{TT}) and the correlation function of
${\cal O}_+({\bf x})$ and ${\cal O}_+({\bf y})$ is in the logarithmic
form (\ref{TTT}).
The constants $c$ and $c '$ which appear in the correlation functions
(\ref{TT}) and (\ref{TTT})
are related to
the parameters $m$ and $M$ appearing in (\ref{ninteen}). This shows that in
the boundary of two layers AdS space,
which is in fact a non-commutative double sheets $d$--
dimensional Minkowski space, we have a quantum field theory.
Also, we note that by using a generalized action for the interacting
scalr field in the two layers AdS space, one naturally get the $n$--point
correlation functions of a quantum field theory in the
$d$--dimensional non-commutative Minkowski spaces \cite{GHEZ}.
\Section{Conclusions}

In this article, we have shown how one can use the approach of non-commutative
geometry to obtain a classical field theory for the spinor and scalar fields
in the classical or commutative spaces (i.e. the spaces which can be
expressed by a commutative algebra). We have used a two layers Anti
de Sitter space
and by the above mentioned approach, we have obtained a classical field
theory for spinors and
scalars in the bulk space. However by considering the AdS/CFT
correspondence, our goal is to correspond the generating functions of these
theories to the quantum correlation functions of a
logarithmic conformal field theory on the boundaries of the
AdS space.
On the other hand, we have presented a geometrical model
for the fermionic and bosonic singleton theories, i.e. equations (4.3)
and (5.3).
In fact, in \cite{GHEZ}, an algebraic model was established for these
theories by taking the derivative of the ordinary Klein-Gordon
and Dirac actions with respect to some parameter.
However, in this article, we have shown that one can obtain the
singleton actions from a generalized Dirac or Klein-Gordon
theory in a non-commutative space.  \newline
This consideration makes an idea for solving the parameters restriction in
Connes-Lott model for the geometrization of standard model. In fact if
someone tries to obtain the Connes-Lott model in the boundary of a two layer
space with the appropriate Higgs potential by considering a suitable theory
in the bulk, then the theory on the boundary is a quantum theory, without
any parameters restriction. This is what we are going to show explicitly in
near future.

\vskip 1cm

{\large {\bf Acknowledgement}}\\
\vspace {.2cm}
The authors would like to thank A.H. Fatollahi and
S. Parvizi for collaboration in the early stage of this
work.

\end{document}